\documentclass[referee,a4paper,12pt,traditabstract]{swsc} 

\usepackage{graphicx}
\usepackage{txfonts}
\usepackage{subfigure}
\usepackage{epstopdf}
\usepackage{lineno}
\usepackage[authoryear,round]{natbib}
\usepackage[backref]{hyperref}
\usepackage{url}

\hypersetup{colorlinks=true,citecolor=cyan,urlcolor=cyan,linkcolor=blue}

\begin{document}

%\begin{linenumbers}

   \title{The dependence of the [FUV-MUV] colour on solar cycle}
   
   \titlerunning{The dependence of the [FUV-MUV] colour on solar cycle}

   \authorrunning{Lovric et al.}

\author {M. Lovric \inst{1},
F. Tosone\inst{1,2},
E. Pietropaolo\inst{2},
D. Del Moro \inst{1},
L. Giovannelli \inst{1},
C. Cagnazzo  \inst{3}
 \and
F. Berrilli \inst{1}
 \fnmsep\thanks{Corresponding author}
 }

    \institute{Department of Physics, University of Rome Tor Vergata,
             Via Ricerca Scientifica,1, I00133, Rome, Italy\\
              \email{francesco.berrillil@roma2.infn.it}
 \and
            Department of Physics and Chemistry, University of L'Aquila, Via Vetoio 10, I67010 Coppito (L'Aquila), Italy
 \and
            Institute of Atmospheric Sciences and Climate (CNR-ISAC),  Via Fosso del Cavaliere, 10, I00133, Rome, Italy\\
            %\thanks{At the moment alphabetic order, but we'll define the final order later}
             }

%%   \date{Received September 15, 1996; accepted March 16, 1997}

  % \abstract{}{}{}{}{}        %% uncomment if structured abstract is desired
 %% 5 {} token are mandatory
 
  \abstract
 %% context heading (optional). leave {} empty if necessary  
   {Solar UV variability  is extremely relevant for the stratospheric ozone. It has an impact on Earth's atmospheric structure and dynamics through radiative heating and ozone photochemistry. Our goal is to study the slope of the solar UV spectrum in two UV bands important to the stratospheric ozone production.
   %}        %% uncomment if structured abstract is selected
 %% aims heading (mandatory)
   %{        %% uncomment if structured abstract is selected
   In order to investigate the solar spectral variability, we use data from SOLSTICE (the Solar Stellar Irradiance Comparison Experiment) onboard the Solar Radiation and Climate Experiment (SORCE) satellite. Data sets used are far UV (115-180nm) and middle UV (180-310nm), as well as the Mg II index (the Bremen composite). 
   %}        %% uncomment if structured abstract is selected
 %% methods heading (mandatory)
   %{        %% uncomment if structured abstract is selected
   We introduce the SOLSTICE $[FUV-MUV]$ colour to study the solar spectral characteristics, as well as to analyse the colour versus Mg II index. To isolate the 11-year scale variation, we used Empirical Mode Decomposition (EMD) on the data sets.
   %}        %% uncomment if structured abstract is selected
 %% results heading (mandatory)
   %{        %% uncomment if structured abstract is selected
     The $[FUV-MUV]$ colour strongly correlates with the Mg II index. The $[FUV-MUV]$ colour shows a time dependent behaviour when plotted versus the Mg II index. To explain this dependence we hypothesize an efficiency reduction of SOLSTICE FUV irradiance using an exponential aging law.  \\
   
   %}        %% uncomment if structured abstract is selected
 %% conclusions heading (optional), leave {} empty if necessary 
   }        %% replace by pair of curly brackets, {}, if structured abstract is selected

   \keywords{sun - solar activity - UV flux - spectral solar irradiance - solar cycle}

   \maketitle
%%
%%________________________________________________________________

\section{Introduction}
  The radiative and particle output of the Sun is variable on different time scales \citep{Hathaway}, from seconds to the evolutionary scale of the star. These fluctuations, due to instabilities and non-stationary processes related to solar magnetic field dynamics, turbulent convection and evolutionary mechanisms, affect the energy balance of the Earth's surface and atmosphere, thus influencing our climate \citep[e.g.][]{Houghton, herman78, Kuhn, Andrews, Foukal, Haigh, Goldbaum, Lockwood}. The main component of the solar variability is the well-known 11-year Schwabe cycle. This cycle can be observed with a number of different characteristics, both physical (e.g., Total Solar Irradiance; \citealp{Kopp14}; Solar Spectral Irradiance \citealp{Floyd2003}; Mg II; \citealp{Heath86, Dudok}; or F10.7 fluxes; \citealp{tapping87, tapping13}; and synthetic: e.g. sunspot number and areas;  \citealp{Hathaway} and references therein). Variations in these indices demonstrate both large and small fluctuations. Whereas the sunspot number index shows large variations in the amplitude of the cycle, Total Solar Irradiance (TSI) only shows changes of around 0.1\% with different spectral regions contributing different amounts (e.g., the UV). The 11-year averaged TSI appears to increase by 0.09\% since the Maunder minimum \citep{Krivova10}.
  
Different parts of the solar spectrum interact with different regions of the Earth's atmosphere, with a non-linear impact on the atmospheric structure that depends on altitude, latitude and season \citep{Meier}. 
The interaction between solar variability and climate is also related to changes in the atmospheric circulation within the stratosphere \citep[e.g.][]{Gray}, where the absorption  by ozone of solar UV $200-–300$ nm photons leads to the radiative heating observed in the stratosphere and to the formation of positive vertical temperature gradients \citep[e.g.][]{bordi}. 
The impact of the large variability in the Solar Spectral Irradiance (SSI) below 400 nm on the terrestrial atmosphere, through radiative heating and ozone photochemistry,
has been recently reviewed in \citet{Ermolli}.

Over the last 13 years, the analysis of the Spectral Irradiance Monitor (SIM) data suggests that the SSI values of the SORCE experiment \citep{Woods} exhibit a brightening for wavelengths with a brightness temperature greater than 5770 K, and a dimming for wavelengths with smaller brightness temperatures, during the solar cycle's declining phase \citep{Harder}. This behavior suggests that SSI variability throughout the 11-year Schwabe cycle could be significantly higher than thought previously, especially in the UV spectral region \citep[e.g.][]{Harder}, which has an important impact on stratospheric ozone \citep{Haigh}. \citet{DeLand04} suggest that SORCE instrumentation correction errors can account for a large part of the reported discrepancy. In addition, \citet{Haigh} noted that spectral variations in the UV spectral region during the solar cycle's declining phase since April 2004, led to a significant decrease in stratospheric ozone, for altitudes less than 45 km, and an increase above that altitude.

As UV radiation is mostly absorbed by the Earth's atmosphere, measurements must be carried out in space. SOLSTICE/UARS (Upper Atmosphere Research Satellite) \citep{Rottman93} and SUSIM (Solar Ultraviolet Spectral Irradiance Monitor) on board UARS measured the UV radiation between 120 and 400 nm, starting in 1991 and ending in 2001 and 2003, respectively. A long record of UV measurements is provided by SOLSTICE on board SORCE, which was launched in 2003.
Unfortunately, instruments age in space resulting in a degradation of the UV signal. Consequently, care must be taken during any analysis of solar UV observations, especially in the merging and correction of data from different space experiments. Recent papers \citep[e.g.] []{DeLand04, lean2012, ball2016} show inconsistency of the SORCE SIM and SOLSTICE data, between the declining phase of solar cycle (SC) 23 and the rising phase of SC 24, with the previous results and with proxy data. The inconsistencies are most likely of instrumental origin. For this reason, we turn to solar proxies to create empirical models that allow us to merge SSI measurements and reconstruct missing observations \citep[e.g.][]{Scholl}. One such proxy is the Mg II core-to-wing ratio \citep[e.g.] []{Heath86, Viereck01}, which is a good indicator of the solar UV irradiance. In addition, there are semi-empirical models that can be used to recover the SSI and TSI signals using different quiet and magnetic surface components automatically  segmented in full-disc magnetograms and continuum images \citep[e.g.][]{Yeo}.

Our approach uses a new way to represent the slope of the solar UV spectrum in the region relevant for stratospheric ozone production. We define the SOLSTICE colour index $[FUV-MUV]$ and study its dependence on the solar 11-year Schwabe cycle using the Mg II index as a proxy for solar activity \citep[e.g.][]{Snow, Viereck99}. The $[FUV-MUV]$ colour index allows empirically-motivated UV reconstructions over the cycles where suitable Mg II data exist. These reconstructions will be useful for addressing the issue of stratospheric ozone dependence on the colour index $[FUV-MUV]$ during the 11-year Schwabe cycle. Moreover, the colour index may be useful for application to observational surveys of UV activity in Sun-like stars.
%_________________________________________________________ sun-like stars observation may be useful for application to sun-like stars observation _______ SSI and MgII DATASETS
\section{Datasets and data preparation}

%________________________________________________________________
\subsection{Magnesium II index dataset}
The Mg II core-to-wing ratio is a good proxy for the solar UV irradiance and facular component of the TSI \citep[e.g.][]{Dudok}.
The Mg II index is calculated as the ratio between the {\it h} and {\it k} emission doublet at 280 nm, and a reference continuum intensity at specific wavelengths in the wings of the Mg II absorption band (see Figure 1 in \citealp{Skupin}). The former originates in the chromosphere, while the latter originates in the photosphere \citep{Bruevich13}. The Mg II index as calculated by \citet{Heath86} as:
\begin{equation} 
I = \frac{4[E_{\mathrm{279.8}}+E_{\mathrm{280.0}}+E_{\mathrm{280.2}}]}{3[E_{\mathrm{276.6}}+E_{\mathrm{276.8}}+E_{\mathrm{283.2}}+E_{\mathrm{283.4}}]} ,
\end{equation}
where $I$ is the Mg II index and $E_\lambda$ is spectral irradiance measured at seven specific wavelengths.

The Mg II index is a ratio of measurements, and as a consequence it is robust against instrument degradation and ageing factors \citep{Skupin}.
The Mg II dataset used for the work presented here is the Bremen composite obtained from SBUV/2 \citep{Viereck01}, UARS SOLSTICE and the GOME instrument on board the ERS-2 satellite from 1995-2011, SCIAMACHY from 2002-2012, GOME-2A from 2007 to today and GOME-2B from 2012 to today. We note there are other instruments taking measurements from which the Mg II index is being derived, although they are not used in our work ( e.g. Mg II derived from SUSIM observations, see \citealp{Brueckner}).

The Mg II Bremen composite, downloaded from the University of Bremen website (\url{http://www.iup.uni-bremen.de/gome/gomemgii.html}), gives daily Mg II index values, from which we calculated monthly mean values. Mg II index data are available since 1978, but in this particular case we only use the data from May 2003 onwards, since the SORCE SOLSTICE data are not available prior to that date.

\subsection{SORCE SOLSTICE dataset}\label{solstice}
The SORCE mission has a scientific payload with different instruments that measure the total solar irradiance, as well as the spectral solar irradiance (in the UV, visible, near infrared and X-ray bands), the same optics and detectors are used to observe other bright stars as well. Solar measurements are performed when the satellite is positioned between the Earth and the Sun, while measurements of other bright stars are performed when the Earth is positioned between the satellite and the Sun \citep{Rottman05}.
SOLSTICE \citep[]{mcclintock} consists of two instruments, SOLSTICE A and SOLSTICE B, which are spectrometers that isolate and detect wavelengths between $115$ nm and $320$ nm daily, with a spectral resolution of 0.1nm, while the spectral resolution of the data set we use here is 1nm. SOLSTICE uses a single optical system for both solar and stellar spectral irradiance observations \citep{msw2005}. It provides one of the longest records of SSI measurements in the UV spectral range, after the SUSIM instrument on board UARS.

The FUV and MUV data sets were downloaded from the LASP (Laboratory for Atmospheric and Space Physics) SORCE website \url{http://lasp.colorado.edu/home/sorce/data/}, and are the version 15 data. The relative accuracy, approximately $0.5\%$ around 310 nm, has a negligible impact on our analysis of [FUV-MUV] colour index and do not influence its slope (see Fig. \ref{FigTempTime}). For more information, see SOLSTICE data product release notes \url{http://lasp.colorado.edu/home/sorce/instruments/solstice/solstice-data-product-release-notes/}.

\section{Data analysis and results}
\subsection{Data analysis}\label{Inter}
Due to the large number of missing data, which leads to non physical gaps in the monthly means of the intensities in both spectral regions, the first step in processing the FUV and MUV daily data was to interpolate. A Piecewise Cubic Hermite Interpolating Polynomial (PCHIP) has been implemented, resulting in the elimination of all the non physical gaps.

After the interpolation, the integration of the irradiance over all wavelengths was performed. For this purpose we use Riemann integration. After obtaining the daily integrated data set, we calculate the monthly mean values, as opposed to the standard $\sim$27 days mean corresponding to the solar rotation, to be compatible with the approach used in geophysics. The calculated monthly mean time series will be decomposed into narrow-band time components in section \ref{EMD}.

%

%__________________________________________________________________ Colour index [FUV-MUV]
\subsection {The SOLSTICE [FUV-MUV] colour index}\label{colour}

The time dependence of the slope of the solar UV spectrum is investigated using the SOLSTICE
[FUV-MUV] colour index. The colour index is a measure of the ratio of the fluxes in two spectral regions \citep{Lena} and is calculated as the difference between the stellar magnitudes at two  wavelengths. The definition of colours with synthetic photometry from spectrophotometry should also be used to gain insights into the spectra of Sun-like stars for future large surveys with next generation space telescopes.

In this work the [FUV-MUV] colour index is calculated using the solar integrated fluxes in the bands FUV (115-180 nm) and MUV (180-310 nm). As usual in stellar photometric systems \citep{Bessel2005} we define a nominal central wavelength of the filter bandpass, known also as the effective wavelength of the band. The central wavelengths of the FUV and MUV bands are 147.5nm and 245nm, respectively. 

The [FUV-MUV] colour index is defined as:
\begin{equation} 
[FUV-MUV] = -2.5 \log \frac{F_{FUV}}{F_{MUV}} + Z_{FUV} - Z_{MUV},
\end{equation} 
where $Z_{FUV}$ and $ Z_{MUV}$ are the zero points of our magnitude scale.
Usually, the zero points are chosen so as to make the colour index of A0 main-sequence star zero \citep{Lena}. In our case they are arbitrarily set to zero corresponding to a offset in magnitude, i.e. a simple shift of all the magnitudes vertically by a constant amount. Therefore, the [FUV-MUV] colour is:
\begin{equation} 
[FUV-MUV] = -2.5 \log \frac{F_{FUV}}{F_{MUV}}.
\end{equation}
%--------------------------------- figure
   \begin{figure}
   \centering
   \includegraphics[width=9cm]{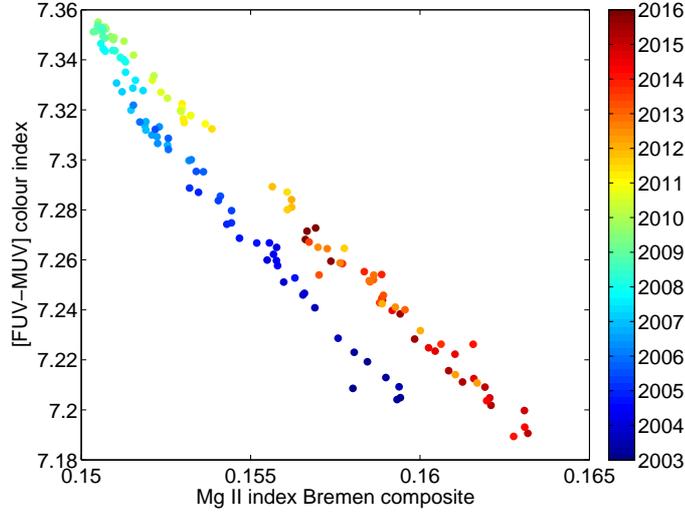}
      \caption{\small The colour index dependence on the Mg II index and time. Date is represented by the colour bar (shown on the right). Data points for November 2012, August, September, October and November 2013 and January 2014 are not shown, as SSI data is not available for these months. Note that the strong correlation between the [FUV-MUV] colour an the Mg II index shows two slightly different slopes, corresponding to the descending phase of solar cycle 23 and to the rising phase of solar cycle 24.
              }
         \label{FigTempTime}
   \end{figure}

Figure \ref{FigTempTime} shows the dependence of the [FUV-MUV] colour on both Mg II index and time and is based on data sets that cover the time between May 2003 (three years after the maximum of cycle 23) and October 2015 (one year and half after the maximum of cycle 24). It is clear that there is a strong correlation between the colour index and Mg II index. This means that the slope of the solar UV spectrum, i.e., the relative variation of considered fluxes, is proportional to solar activity on the time scale of 11 years, signifying a stronger growth of FUV flux than MUV flux when the solar magnetic activity increases. In other words, the colour index decreases as solar activity increases.

Furthermore, the time dependence of the [FUV-MUV] colour points out the existence of two slightly different slopes, corresponding to the descending phase of solar cycle 23 and to the rising phase of solar cycle 24. Starting from the minimum in [FUV-MUV] colour corresponding to May 2003, the Mg II index is seen to decrease while the colour index increases.\\ 
Around the maximum in 2013, the Mg II index reaches a maximum of $0.163$ and the colour index is at the minimum value of $7.19$. From this point, until the date of last observation, the colour index dependence on Mg II index is retracing the same path in parameter space in reverse.

In principle, the two different slopes can be explained by the dependence of the spectral properties of magnetic structures \citep[e.g.][]{Meunier} on the cycle of solar activity, or by a residual uncorrected SOLSTICE instrument degradation \citep[e.g.][]{Snow}.

Although unlikely, the solar spectral energy distribution (SED) can be modified during the cycle by various physical effects: {\it i)} the variable photometric behavior of a star that is faculae-dominated like the Sun \citep[e.g.][]{LockwoodGW}, {\it ii)} the possible presence of North-South asymmetries of activity \citep[e.g.][]{Meunier}, {\it iii)} the presence of an out-of-phase variation between spectral solar irradiance and the Mg II index \citep[e.g.][]{Harder,Li}, and {\it iv)} a center versus limb distance effect, i.e., the solar active regions appear statistically more inclined with respect to the line-of-sight during the SC rising phase compared to the SC descending phase \citep[e.g.][]{Maunder}. Further speculation on this issue seems premature, both because a more detailed analysis of FUV and MUV spectra over a more extended time period is needed, and because we support a model of instrument degradation (see Sec \ref{aging}).\\

\begin{figure}
 \hfill
\subfigure[\small]{\includegraphics[width=8.2cm]{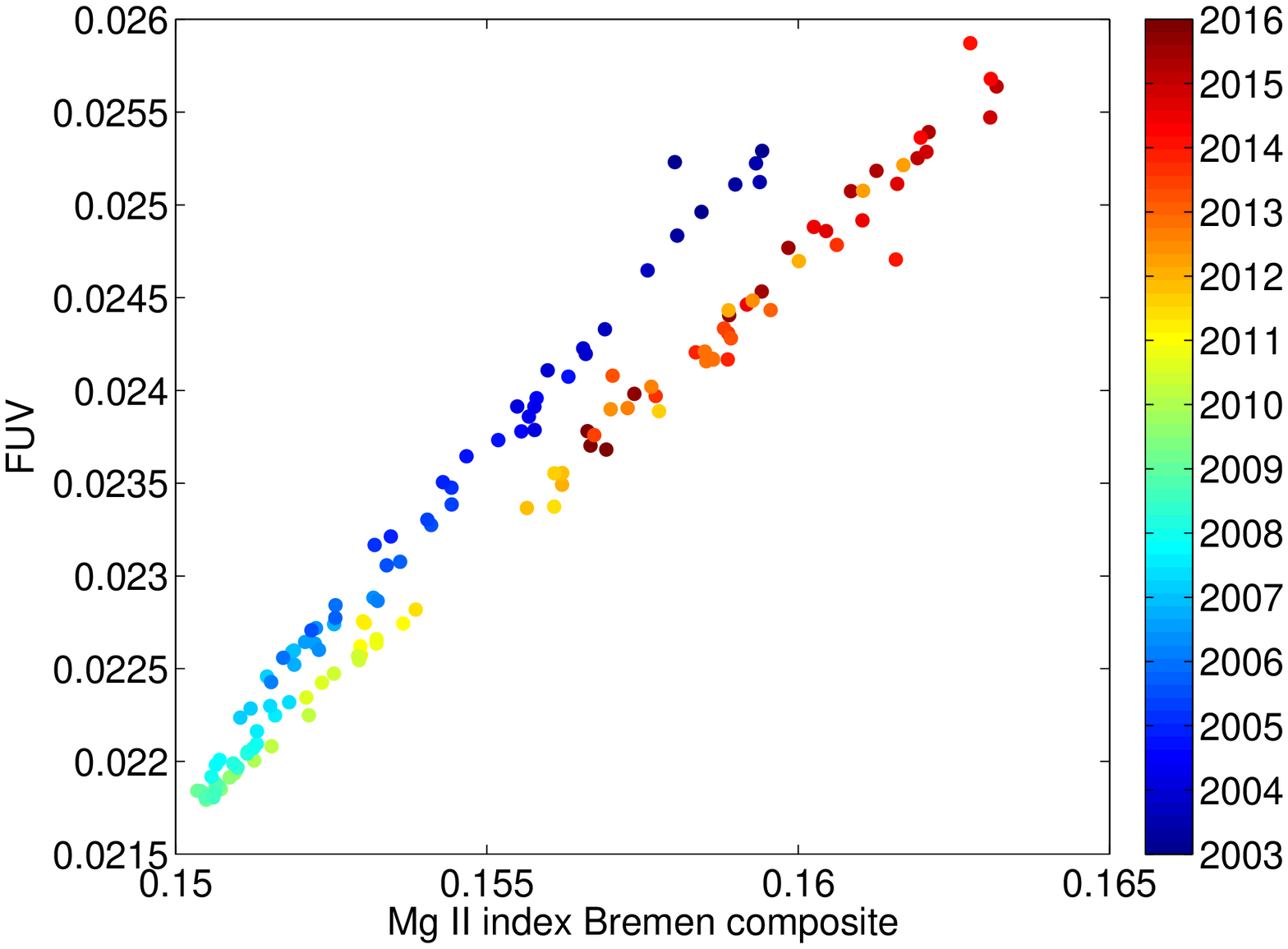}}\label{FUVmgII}
   \hfill
\subfigure[\small]{\includegraphics[width=8cm]{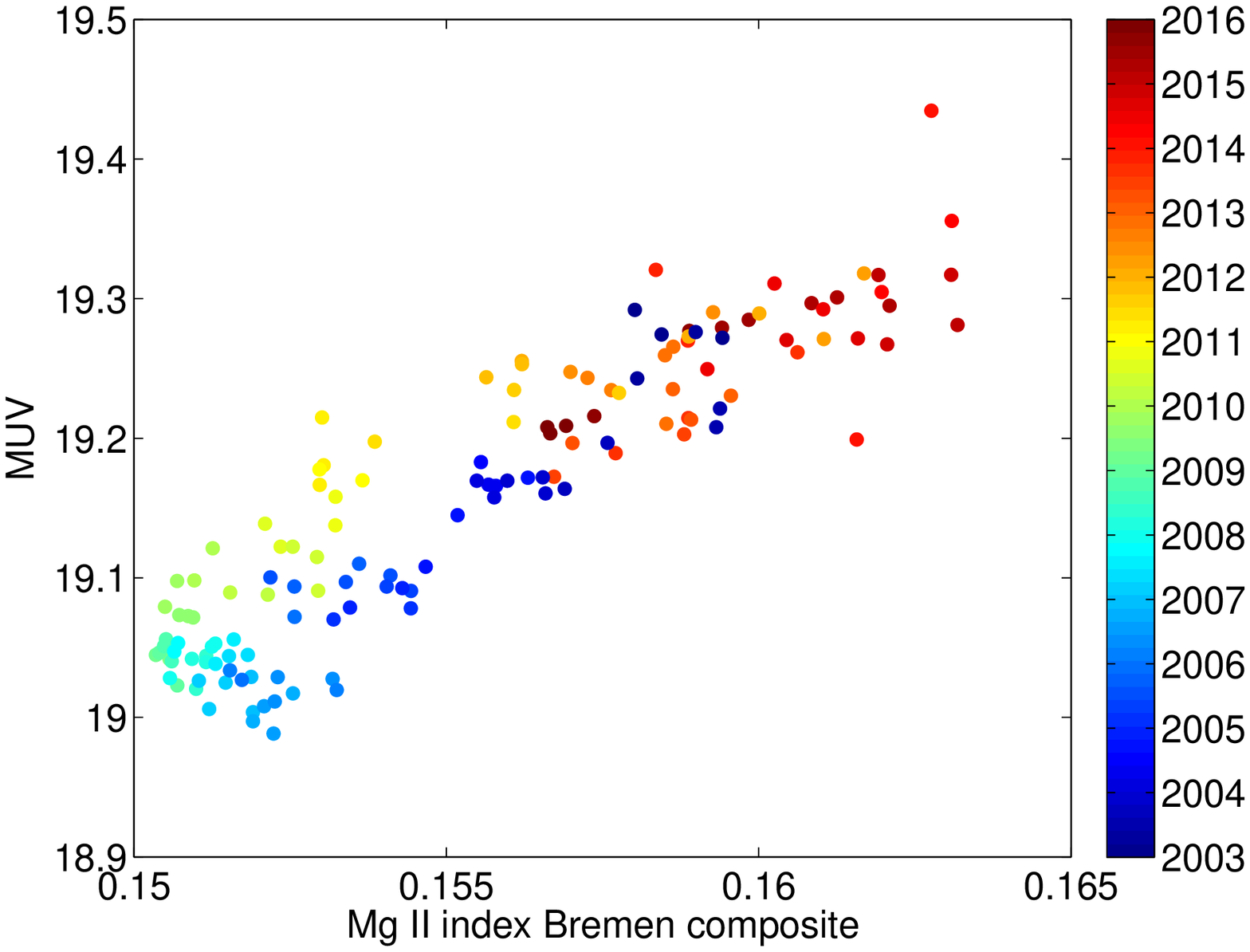}}\label{MUVmgII}
   \hfill
   \caption{\small The FUV $(a)$ and MUV $(b)$ dependence on the Mg II index and time. The effect seen in figure \ref{FigTempTime} arises from the FUV (fig. $(a)$), as two distinct slopes with little scatter are clearly visible, whereas figure $(b)$ exhibits a much larger scatter in the data. This difference can be due to an increased aging effect mainly in the FUV where the energetic EUV radiation alters the instrumental efficiency more than in the MUV spectral region.} \label{FUVMUV} 
   \end{figure}

In order to simulate the variability of the [FUV-MUV] colour index we introduce the [FUV-MUV] colour temperature of the Sun, i.e., the temperature of the blackbody with the same spectral slope in a given wavelength interval \citep{Lena}. In our case the colour temperature is the blackbody temperature required to produce the measured [FUV-MUV] colour:

\begin{equation} 
[FUV-MUV] = -2.5 \log \frac{\lambda_{\mathrm{MUV}}}{\lambda_{\mathrm{FUV}}} \frac{e^ \frac{hc}{\lambda_{\mathrm{MUV}}k_BT_C}-1}{e^ \frac{hc}{\lambda_{\mathrm{FUV}}k_BT_C}-1},
\end{equation}
Where $\lambda_{MUV}$ is average wavelength for the MUV band (245nm), $\lambda_{FUV}$ is average wavelength for the FUV band (147.5nm), c is the speed of light (c = $2.99792458\times10^{8}$ ms$^{-1}$), h is the Planck constant (h = $6.6260755\times10^{-34}$ Js), $k_{B}$ is the Boltzmann constant ($k_{B}$ = $1.380658\times10^{-23}$ JK$^{-1}$), and
$T_C$ is the Colour temperature. \\

It is important to note that a blackbody radiator is not a realistic representation of the solar atmosphere in the UV spectral region. The rapid increase in temperature that occurs in the transition region between the chromosphere and corona produces strong emission lines in the FUV and EUV regions and consequently the solar UV spectrum shows a rich spectrum that can be not represented by the Planck function \citep[e.g.][]{Fontenla}.
Nevertheless, we find that the observed [FUV-MUV] colour variation is consistent with a variation of the [FUV-MUV] colour temperature of about 100 K between the minimum of cycle 23 ($\simeq 5340$ K) and the maximum of cycle 24 ($\simeq 5440$  K). We will use this result in Section \ref{aging} to model a residual degradation effect in the SOLSTICE data.

%EM
%=======================================================================
\subsection{Empirical  Mode  Decomposition analysis}\label{EMD}
We use the Empirical  Mode  Decomposition (EMD) technique to investigate the dependence of the $[FUV-MUV]$ colour index and Mg II signal at various time scales, particularly at the 11-year scale. The complexity of the physics associated with the generation of solar spectral variability makes these signals intrinsically non-stationary and nonlinear, therefore an adaptive analysis is required, i.e., where the decomposition basis is derived from the data \citep{Huang}. The EMD analysis is a signal processing technique designed to analyse such kind of signals. We refer to \citet{Huang} for a detailed description of the method. The EMD's main aim is to synthesize any signal as the sum of a finite number of Intrinsic Mode Functions (IMFs) computed directly from the signal. The IMFs generated with this process satisfy two conditions: 

\begin{enumerate}
\item The number of zeros and of extrema differs at most by one;
\item The mean value of the envelopes of maxima and minima is approximately zero.
\end{enumerate}

The IMFs result in quasi symmetric functions with respect to their mean value, and are representative of a mode or a trend embedded in the signal. They are not required to be harmonic, and are suitable for the derivation of signals with anharmonic form, such as the solar signals.

The IMF functions are extracted with an iterative process that stops when a threshold is reached;  Huang's original criterion takes into account the quadratic relative error computed between the signal components after two consecutive iterations, stating that the sum over the whole data set should be less than $30\%$. This stopping criterion is not adaptive since it does not take into account the possibility of a slowly varying standard deviation.

\begin{figure}
 \centering
  \includegraphics[width=15cm]{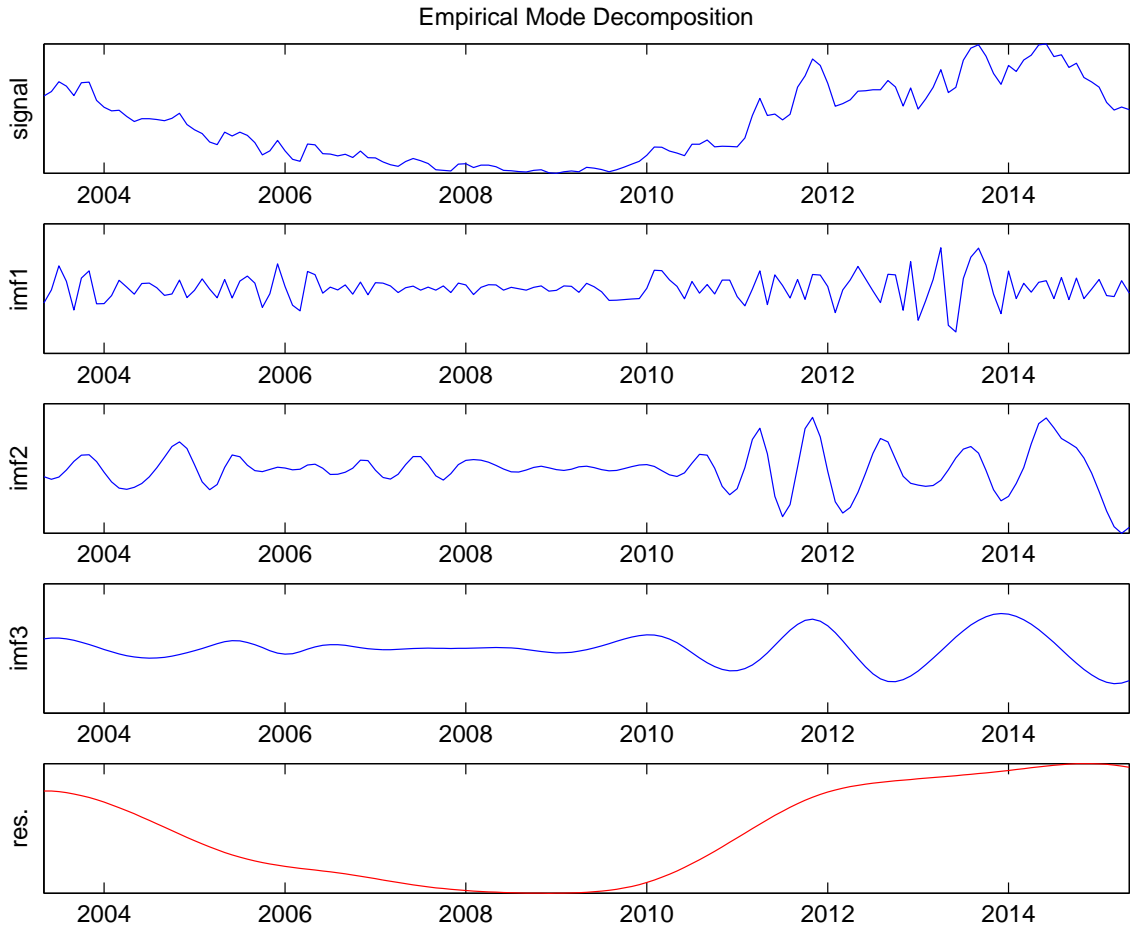}
  \caption{\small The IMFs of the Mg II index are shown from the first (highest frequency) to the last (lowest frequency) starting from the top. The data sets cover the time between May 2003 and October 2015. The bottom panel corresponds to the 11-year Schwabe cycle.}
         \label{mg_imfs}
\end{figure}

\begin{figure}
 \centering
  \includegraphics[width=15cm]{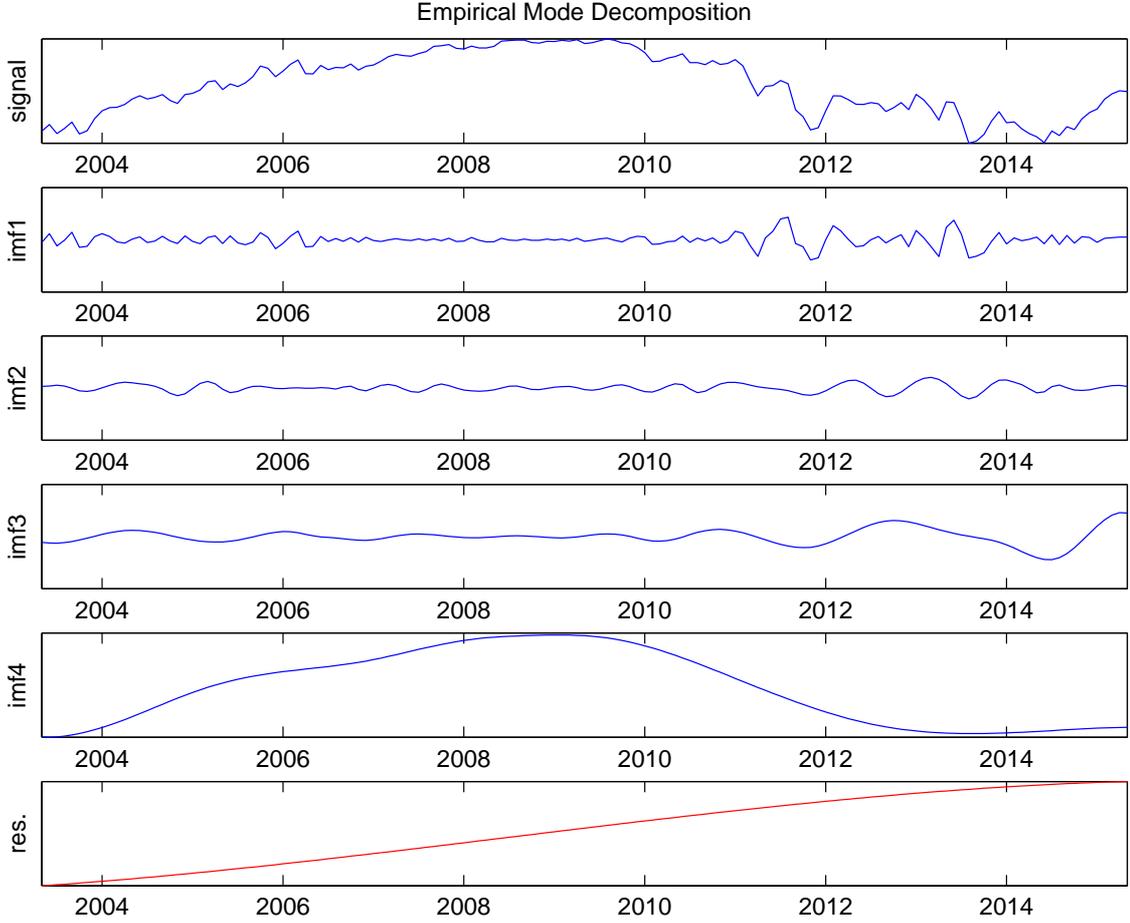}
   \caption{\small The IMFs of the colour index $[FUV-MUV]$ are shown from the first (highest frequency) to the last (lowest frequency) starting from the top. The data sets cover the time between May 2003 and October 2015. IMF4 corresponds to the 11-year Schwabe cycle, while the bottom panel, labeled "res.", represents a residual trend  most likely due to aging effects on SOLSTICE data.}
         \label{colour_imfs}
\end{figure}

In order to overcome this limitation, we apply an adaptive stopping criterion (for more information see section 3.2 in \citet{Rilling}, to extract the IMFs representative of the solar signals. We extract 4 and 5 different functions for the Mg II index and $[FUV-MUV]$ colour, respectively. As an example, we show in figure \ref{colour_imfs} the IMFs of the $[FUV-MUV]$ colour.\\
For both signals, the first two IMFs are associated with the "solar noise". These IMFs contain the signal associated with the solar rotation and with the evolution of large active regions, that are responsible for the medium time scale variation of solar UV spectral signal, i.e., a few months.
A quasi biennial variation \citep[e.g.][]{Penza06} dominates the third IMF. This finding requires future study to investigate the role of SSI quasi biennial variations in the variability of stratospheric ozone. \\
It is worth noting that the EMD analysis generates the 11-year period and trend components of the signal in a different way for the $[FUV-MUV]$ colour index and Mg II signal. In particular, in order to synthesise the 11-years period and trend components of the $[FUV-MUV]$ colour index it is necessary to use the IMF4 and the residual (labeled as "res", see figure \ref{colour_imfs}), representing a residual trend most likely due to aging effects on SOLSTICE data. For the Mg II signal the residual function is sufficient to synthesise the 11-years period. All functions shown in figures \ref{mg_imfs} and \ref{colour_imfs} are normalized to show their general behaviour \citep{Rilling}.

\begin{figure}
 \centering
  \includegraphics[width=10cm]{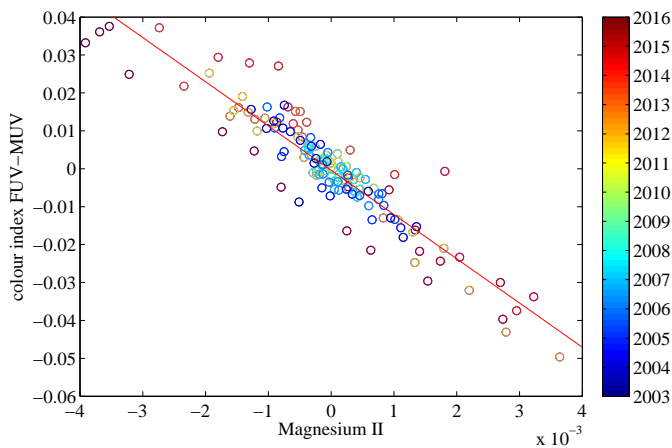}
   \caption{\small Scatter plot between the $[FUV-MUV]$ colour index (normalized arbitrary units) and Mg II index (normalized arbitrary units) reconstructed after removing the IMFs representing the 11-year signal. The correlation implies that even at the shortest temporal scales the FUV flux variability dominates the MUV one when the solar activity increases. Time is represented as in Figure \ref{FigTempTime}}
         \label{ScatterNo11}
\end{figure}

Moreover, IMFs are used to investigate the dependence of the $[FUV-MUV]$ colour index versus Mg II correlation with short time scale components of the solar cycle. (Fig. \ref{ScatterNo11}) shows the $[FUV-MUV]$ colour index versus Mg II correlation reconstructed using the first three IMFs, not associated with the 11-year component. This correlation is especially interesting because it implies that
as the solar activity increases, the $[FUV-MUV]$ colour index decreases, also using the first EMD components containing the shortest temporal scales.

%__________________________________________________________________ SORCE aging MODEL
\section{SOLSTICE aging model}\label{aging}
Although the SOLSTICE experiment is designed to strongly reduce the possible effect of detector aging (see section \ref{solstice}), a residual degradation can explain the presence of two different slopes in the colour-Mg II correlation during the descending and rising phases of the solar cycle, and the reverse path of the last two years (see Fig. \ref{FigTempTime}). 

To simulate the colour-dependent index changing over the period used for our analysis, we hypothesize a residual non-corrected degradation in the FUV irradiance measurements as the result of the aging of the optics or the detection process due to EUV radiation damage. The assumed efficiency degradation in the FUV spectral region is integrated over the whole FUV bandpasses, i.e., we assume the same degradation for all wavelengths in the FUV spectral region. The calculated $[FUV-MUV]$ colour temperature variation of the Sun during the analysed period (see section \ref{colour}) is used to simulate the observed variability of the $[FUV-MUV]$ colour. 

\begin{figure}
   \centering
   \includegraphics[width=10cm]{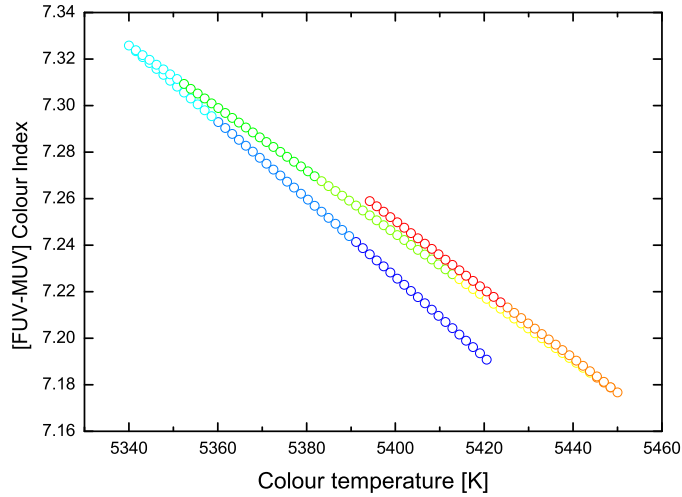}
      \caption{\small The modeled colour index dependence on colour temperature. Coloured symbols show the colour index variation assuming a FUV channel aging described by an exponential law with a degradation constant $D = 0.04$ and an effective degradation lifetime $\tau$=70 months. Deep blue corresponds to 2003, while deep red corresponds to 2015.
              }
         \label{FigVibStab}
   \end{figure}

More in detail, to reproduce such a behaviour (see Figure \ref{FigTempTime}), we hypothesize a linear time dependence of the $[FUV-MUV]$ colour index during the solar cycle and an exponential degradation of optics or detection processes in the FUV spectral region. The variability of the $[FUV-MUV]$ colour is simulated with the estimated variation of about 100 K between the colour temperature associated with the minimum of cycle 23 ($T_C \simeq 5340$ K) and the colour temperature associated  with the maximum of cycle 24 ($T_C \simeq 5440$  K). Thus we hypothesize an aging exponential reduction of FUV irradiance using:
\begin{equation} 
E_{age}(t) = 1-D e^{-\frac{t}{\tau}}
\end{equation}
where $E_{age}(t)$ represents the deviations from the reference in the FUV spectral region, $D$ is a degradation constant, $t$ the time expressed in months, and $\tau$ the effective degradation lifetime in months.
The two parameters of this aging law $E_{age}(t)$, that closely approximates the experimentally observed $[FUV-MUV]$ colour dependence on time and Mg II index, are a degradation constant $D = 0.04$ and an effective degradation lifetime $\tau$=70 months. Figure \ref{FigVibStab} shows the modeled behaviour of the colour index as a function of colour temperature and demonstrates that the model reproduces quantitatively the observed features of the $[FUV-MUV]$ colour dependence on time and Mg II index.

The calculated parameters correspond to an average efficiency reduction of a factor $\simeq 0.0002$ per month in the FUV and to a total reduction of about 4\% during the observed period. The corresponding variation of the colour temperature change of about 20 K.

%______________________________________________________________ CONCLUSIONS
\section{Conclusions}

In this paper we address the issue of the relationship between $[FUV-MUV]$ colour index and Mg II index in order to derive the slope of the solar UV spectrum in the region relevant for stratospheric ozone production as a function of the solar activity during the Schwabe cycle. The $[FUV-MUV]$ colour index allows empirically-motivated UV reconstructions over the cycles where suitable Mg II data exist. The colour index may be useful for application to observational surveys of UV activity in sun-like stars.\\

Listed below are the main results obtained from our analysis:
\begin{itemize}
\item We find that the $[FUV-MUV]$ colour index strongly correlates with the Mg II index, with the colour index decreasing as solar activity increases.
This correlation, first reported in \citet{Lovric}, enables us to determine the ratio of fluxes in the FUV and MUV spectral bands, for the period where Mg II index data are available. This provides a complementary approach to the spectral reconstruction models that are typically used \citep[e.g.][]{Thuillier}. A suitable reconstruction of SSI variability, mainly in the FUV and MUV bands, is required to asses the stratospheric ozone variability with solar activity and, consequently, the “top-down” mechanisms capable of amplifying UV solar forcing on the earth's climate.

\item The time dependence of the [FUV-MUV] colour points out the existence of two slightly different slopes, corresponding to the descending phase of solar cycle 23 and to the rising phase of solar cycle 24. Starting from this minimum in [FUV-MUV] colour, the Mg II index is seen to increase while the colour index decreases. From this point, until the date of the last observations, the colour index dependence on the Mg II index retraces the same path in parameter space in reverse. The measured different slopes can be explained by the dependence of the spectral properties of magnetic structures on the cycle of solar activity or by a residual uncorrected SOLSTICE instrument degradation. Our results suggest the latter.

\item We hypothesize a reduction of the instrument efficiency in measuring the FUV irradiance and model it by using an exponential aging.
The two parameters of the exponential aging law $E_{Age}(t)$, that closely approximates the experimentally observed $[FUV-MUV]$ colour dependence on time and Mg II index, are a degradation constant $D = 0.04$ and an effective degradation lifetime $\tau$=70 months (Fig. \ref{FigVibStab}). This corresponds to an average efficiency reduction of a factor $\simeq 0.0002$ per month in the FUV range. This value corresponds to a total reduction of about 4\% during the observed period.

\item We separate temporal intrinsic solar components (i.e., IMFs) in the FUV, MUV and Mg II signals using the EMD technique.
The analysis of $[FUV-MUV]$ colour index versus Mg II index using IMF decomposition supports our conclusion that the two slightly different slopes in the [FUV-MUV] colour  have their origin in the 11-year component and are due to ageing of SOLSTICE FUV. 
IMFs are also used to investigate the dependence of the $[FUV-MUV]$ colour index versus Mg II correlation with short time scale components of the solar cycle. The $[FUV-MUV]$ colour index versus Mg II correlation reconstructed using the first three IMFs not associated with the 11-year component, implies that
as the solar activity increases, the $[FUV-MUV]$ colour index decreases as well, even when using the first EMD components containing the shortest temporal scales.
Moreover, the EMD analysis shows that a quasi biennial variation  dominates the third IMF associated with the $[FUV-MUV]$ colour and Mg II indices. This finding will require future study to investigate the role of solar spectral irradiance fluctuations at quasi biennial scales in the variability of stratospheric ozone.
\end{itemize}

\begin{acknowledgements}
We would like to thank the referees, Jeff Morrill and Martin Snow for careful consideration of the paper and for comments that greatly improved the manuscript. We also thank Prof. Stuart Jefferies and Dr. Roberto Piazzesi for editing the language. We acknowledge G. Rilling for the EMD code (\url{http://perso.ens-lyon.fr/patrick.flandrin/emd.html}). This work was partially supported by the Italian MIUR-PRIN 2012 {\it The active sun and its effects on space and Earth climate} and the Joint Research PhD Program in "Astronomy, Astrophysics and Space Science" between the universities of  Roma Tor Vergata, Roma Sapienza and INAF. ML acknowledges the financial support under AstroMundus, an Erasmus+: Erasmus Mundus Joint Masters Degree program in Astronomy and Astrophysics between the universities of Innsbruck, Belgrade, G\"ottingen, Padua and Roma Tor Vergata. The following institutes are acknowledged for providing the data: Laboratory for Atmospheric and Space Physics (Boulder, Colorado) for SORCE SOLSTICE SSI data (\url{http://lasp.colorado.edu/home/sorce/data/}) and University of Bremen (Bremen, Germany) for Mg II index data (\url{http://www.iup.uni-bremen.de/gome/gomemgii.html}). The editor thanks Martin Snow and Jeff Morrill for their assistance in evaluating this paper.
\end{acknowledgements}

%%    This version assumes use of bibtex with the swsc.bib file being present
%%    If your bib file has a different name you need to change the following line

%\bibliography{swsc}

\begin{thebibliography}{}

\bibitem[Andrews(2000)]{Andrews} Andrews, D.G., An Introduction to Atmospheric Physics, {\it Cambridge University Press, Cambridge, U.K; New York, U.S.A.}, 2000.

\bibitem[Ball et al.(2016)]{ball2016} Ball, W.T., J.D. Haigh, E.V. Rozanov, A. Kuchar, T. Sukhodolov, F. Tummon, A.V. Shapiro, and W. Schmutz, High solar cycle spectral variations inconsistent with stratospheric ozone observations, {\it Nature Geoscience}, {\bf 9}, 206-209, 2016, DOI:10.1038/ngeo2640.

\bibitem[Bessel(2005)]{Bessel2005} Bessel, M. S., Standard Photometric Systems, {\it Annual Review of Astronomy and Astrophysics}, {\bf 43}, 293–336, 2005, DOI:10.1146/annurev.astro.41.082801.100251.

\bibitem[Bordi et al.(2015)]{bordi} Bordi, I., F. Berrilli, and E. Pietropaolo, Long-term response of stratospheric ozone and temperature to solar variability, {\it Annales Geophysicae}, {\bf 33}, 267, 2015, DOI:10.5194/angeo-33-267-2015.

\bibitem[Brueckner et al.(1993)]{Brueckner} Brueckner, G.E., K.L. Edlow, L.E. Floyd, J.L. Lean, and M.E. VanHoosier, The Solar Ultraviolet Spectral Irradiance Monitor (SUSIM) Experiment On Board the Upper Atmosphere Research Satellite (UARS) , {\it Journal of Geophysical Research}, {\bf 98}, 10695, 1993, DOI:10.1029/93JD00410.

\bibitem[Bruevich \& Yakunina(2013)]{Bruevich13} Bruevich, E.A., and G.V. Yakunina, Correlation study of six solar activity indices in the cycles 21 - 23, {\it Sun and Geosphere}, {\bf 8}, 83-90, 2013.

\bibitem[DeLand \& Cebula(2012)]{DeLand04} DeLand M. T. and R. P. Cebula, Solar UV variations during the decline of Cycle 23, {\it Journal of Atmospheric and Solar-Terrestrial Physics}, {\bf 77}, 77, 225-234, 2012, DOI:10.1016/j.jastp.2012.01.007.

\bibitem[Dudok de Wit et al.(2009)]{Dudok}Dudok de Wit, T., M. Kretzschmar, J. Lilensten, and T. Woods, Finding the best proxies for the solar UV irradiance, {\it Geophysical Research Letters}, {\bf 36}, L10107, 2009, DOI:10.1029/2009GL037825.

\bibitem[Ermolli et al.(2013)]{Ermolli} Ermolli, I., K. Matthes, T. Dudok de Wit, N.A. Krivova, K. Tourpali, et al., Recent variability of the solar spectral irradiance and its impact on climate modelling, {\it Atmospheric Chemistry and Physics}, {\bf 13}, 3945-3977, 2013, DOI:10.5194/acp-13-3945-2013.

\bibitem[Floyd et al.(2003)]{Floyd2003} Floyd, L.E., J.W. Cook, L.C. Herring, and P.C. Crane, SUSIM’S 11-year observational record of the solar UV irradiance, {\it Advances in Space Research}, {\bf 31}, 2111–2120, 2003, DOI:10.1016/S0273-1177(03)00148-0.

\bibitem[Fontenla et al.(2011)]{Fontenla} Fontenla, J. M., J. Harder, W. Livingston, M. Snow, and T. Woods, High-resolution solar spectral irradiance from extreme ultraviolet to far infrared, {\it Journal of Geophysical Research}, {\bf 116}, Issue D20, CiteID D20108, 2011, DOI:10.1029/2011JD016032.

\bibitem[Foukal et al.(2006)]{Foukal} Foukal, P., C. Fr\"{o}hlich, H. Spruit, and T.M.L. Wigley, Variations in solar luminosity and their effect on the Earth's climate, {\it Nature}, {\bf 443}, 161-166, 2006, DOI:10.1038/nature05072.

\bibitem[Goldbaum et al.(2009)]{Goldbaum} Goldbaum, N., M.P. Rast, I. Ermolli, J.S. Sands, and F. Berrilli, The Intensity Profile of the Solar Supergranulation, {\it The Astrophysical Journal}, {\bf 707}, 67-73, 2009, DOI:10.1088/0004-637X/707/1/67.

\bibitem[Gray et al.(2010)]{Gray} Gray, L. J., J. Beer, M. Geller, J.D. Haigh, M. Lockwood, et al., Solar Influences on Climate, {\it Reviews of Geophysics}, {\bf 48}, RG4001, 2010, DOI:10.1029/2009RG000282.

\bibitem[Haigh(2007)]{Haigh} Haigh, J.D.,The Sun and the Earth’s Climate, {\it Living Reviews in Solar Physics}, {\bf 4}, A00, 2007, DOI: 10.12942/lrsp-2007-2,
\url{http://citeseerx.ist.psu.edu/viewdoc/download?doi=10.1.1.364.790&rep=rep1&type=pdf}.

\bibitem[Harder et al.(2009)]{Harder} Harder, J. W., J.M. Fontenla, P. Pilewskie, E.C. Richard, and T.N. Woods, Trends in solar spectral irradiance variability in the visible and infrared,  {\it Geophysical Research Letters}, {\bf 36}, L07801, 2009, DOI:10.1029/2008GL036797.

\bibitem[Hathaway(2010)]{Hathaway}Hathaway, D.H.,The Solar Cycle, {\it Living Reviews in Solar Physics}, {\bf 7}, A00, 2010, DOI: 10.12942/lrsp-2010-1
\url{http://www.livingreviews.org/lrsp-2010-1}.

\bibitem[Heath \& Schlesinger(1986)]{Heath86} Heath, D.F., and B.M. Schlesinger, The Mg 280-nm Doublet as a Monitor of Changes in Solar Ultraviolet Irradiance, {\it Journal of Geophysical Research}, {\bf 91}, 8672-8682, 1986, DOI:10.1029/JD091iD08p08672.

\bibitem[Herman(1978)]{herman78}Herman, J.R., and R.A. Goldberg, Sun, weather and climate, {\it NASA Special Publications, 426}, 1978.

\bibitem[Houghton(1977)]{Houghton} Houghton, J.T., The Physics of Atmospheres, {\it Cambridge University Press, Cambridge, U.K; New York, U.S.A.}, 1977.

\bibitem[Huang et al.(1998)]{Huang}Huang N.E., Z. Shen, S.R. Long, M.C. Wu, H.H. Shih, Q. Zheng, N. Yen, C. Tung, and H.H. Liu, The empirical mode decomposition and the Hilbert spectrum for nonlinear and non-stationary time series analysis, {\it Proceedings of the Royal Society A}, {\bf 454}, 903–95, 1998, DOI: 10.1098/rspa.1998.0193.

\bibitem[Kopp(2014)]{Kopp14}Kopp G., An assessment of the solar irradiance record for climate studies, {\it Journal of Space Weather and Space Climate}, {\bf 4}, A14, 2014, DOI:10.1051/swsc/2014012.

\bibitem[Krivova et al.(2010)]{Krivova10} Krivova, N.A., L.E.A. Vieira, and S.K. Solanki, Reconstruction of solar spectral irradiance since the Maunder minimum, {\it Journal of Geophysical Research: Space Physics}, {\bf 115},  A12112, 2010, DOI:10.1029/2010JA015431.

\bibitem[Kuhn et al.(1988)]{Kuhn} Kuhn, J. R., K.G. Libbrecht, and R.H. Dicke, The surface temperature of the sun and changes in the solar constant, {\it Science}, {\bf 242}, 908-911, 1988, DOI:10.1126/science.242.4880.908.

\bibitem[Lean \& DeLand(2012)]{lean2012} Lean, J., and M. DeLand, How does the Sun's spectrum vary?, {\it Journal of Climate}, {\bf 25}, 7, 2555-2560, 2012, DOI:10.1175/JCLI-D-11-00571.1.

\bibitem[L\'ena(1988)]{Lena} L\'ena, P., Observational Astrophysics, {\it Astronomy and Astrophysics Library, Springer-Verlag Eds}, 1988, ISBN 3-540-18433-3.

\bibitem[Li et al.(2016)]{Li} Li, K.J., J.C. Xu, N.B. Xiang, and W. Feng, Phase relations between total solar irradiance and the Mg II index, {\it Advances in Space Research}, {\bf 57}, 408–417, 2016, DOI:http://dx.doi.org/10.1016/j.asr.2015.10.020.

\bibitem[Lockwood(2012)]{Lockwood} Lockwood, M., Solar Influence on Global and Regional Climates, {\it Surveys in Geophysics}, {\bf 33}, 503-534, 2012, DOI:10.1007/s10712-012-9181-3.

\bibitem[Lockwood et al.(2007)]{LockwoodGW} Lockwood, G. W., B.A. Skiff, G.W. Henry, S. Henry, R.R. Radick, S.L. Baliunas, R.A. Donahue, and W. Soon, Patterns of Photometric and Chromospheric Variation among Sun-like Stars: A 20 Year Perspective, {\it The Astrophysical Journal Supplement Series}, {\bf 171}, 260, 2007, DOI:10.1086/516752.

\bibitem[Lovric(2015)]{Lovric} Lovric, M., Solar cycle effects in stratospheric ozone, Master thesis, University of Innsbruck, 2015.

\bibitem[Maunder(1904)]{Maunder}Maunder, E. W.,  Note on the distribution of sun-spots in heliographic latitude, 1874-1902, {\it Monthly Notices of the Royal Astronomical Society}, {\bf 64}, 747-761, 1904.

\bibitem[McClintock et al.(2005)]{mcclintock}  McClintock, W.E.; G.J. Rottman, and T.N. Woods, Solar-Stellar Irradiance Comparison Experiment II (SOLSTICE II): Instrument Concept and Design, {\it Solar Physics}, {\bf 230}, 225-258, 2005.

\bibitem[McClintock et al.(2005)]{msw2005}  McClintock, W.E., M. Snow, and T.N. Woods, Solar–Stellar Irradiance Comparison Experiment II (SOLSTICE II): Pre-Launch and On-Orbit Calibrations, {\it Solar Physics}, {\bf 230}, 259-294, 2005.

\bibitem[Meier(1991)]{Meier} Meier, R. R., Ultraviolet spectroscopy and remote sensing of the upper atmosphere, {\it Space Science Reviews}, {\bf 58}, 1-185, 1991, DOI:10.1007/BF01206000.

\bibitem[Meunier(2003)]{Meunier} Meunier, N., Statistical properties of magnetic structures: Their dependence on scale and solar activity,  {\it Astronomy \& Astrophysics}, {\bf 405}, 1107, 2003, DOI:0004-6361:20030713.

\bibitem[Penza et al.(2006)]{Penza06} Penza, V., E. Pietropaolo, and W. Livingston, Modeling the cyclic modulation of photospheric lines, {\it Astronomy \& Astrophysics}, {\bf 454}, 349-358, 2006, DOI:10.1051/0004-6361:20053405.

\bibitem[Rilling et al.(2003)]{Rilling}Rilling, G., P. Flandrin, and P. Gon\c{c}alves, On empirical mode decomposition and its algorithms, {\it IEEE-EURASIP workshop on nonlinear signal and image processing}, {\bf 3}, NSIP-03, Grado (I), 2003.

\bibitem[Rottman et al.(1993)]{Rottman93} Rottman, G.J., T.N. Woods, and T.P. Sparn, Solar-Stellar Irradiance Comparison Experiment 1. I - Instrument design and operation,  {\it Journal of Geophysical Research}, {\bf 98}, 10, 667, 1993, DOI:10.1029/93JD00462.

\bibitem[Rottman(2005)]{Rottman05} Rottman, G.J., The SORCE Mission, {\it Solar Physics}, {\bf 230}, 7-25, 2005.

\bibitem[Sch\"oll et al.(2016)]{Scholl} Sch\"oll, M., T. Dudok de Wit, M. Kretzschmar, and M. Haberreiter, Making of a solar spectral irradiance dataset I: observations, uncertainties, and methods, {\it Journal of Space Weather and Space Climate}, {\bf 6}, A14, 2016, DOI:http://dx.doi.org/10.1051/swsc/2016007.

\bibitem[Skupin et al.(2005)]{Skupin} Skupin, J., S. Noyel, M.W. Wuttke, M. Gottwald, H. Bovensmann, M. Weber, and J.P. Burrows, GOME and SCIAMACHY solar spectral irradiance and Mg II solar activity proxy indicator,  {\it Memorie della Societa Astronomica Italiana}, {\bf 76}, 1038-1041, 2005.

\bibitem[Snow et al.(2014)]{Snow} Snow, M., M. Weber, J. Machol, R. Viereck, and E. Richard, Comparison of Magnesium II core-to-wing ratio observations during solar minimum 23/24, {\it Journal of Space Weather and Space Climate}, {\bf 28}, A04, 2014, DOI:10.1051/swsc/2014001.

\bibitem[Tapping(2013)]{tapping13} Tapping, K. F., The 10.7 cm solar radio flux ($F_{10.7}$), {\it Space Weather}, {\bf 11}, 394–406, 2013, DOI:10.1002/swe.20064.

\bibitem[Tapping(1987)]{tapping87} Tapping, K. F.,  Recent  solar  radio  astronomy  at  centimeter wavelengths: The variability of the 10.7 cm flux, {\it Journal of Geophysical Research}, {\bf 92}, 829–838, 1987, DOI:10.1029/JD092iD01p00829.

\bibitem[Thuillier et al.(2012)]{Thuillier} Thuillier, G., M. Deland, A. Shapiro, W. Schmutz, D. Bolsée, and S.M.L. Melo, The Solar Spectral Irradiance as a Function of the Mg ii Index for Atmosphere and Climate Modelling, {\it Solar Physics}, {\bf 277}, 245, 2012, DOI:10.1007/s11207-011-9912-5.

\bibitem[Viereck \& Puga(1999)]{Viereck99} Viereck, R., and L. Puga, The NOAA Mg II core-to-wing solar index: Construction of a 20-year time series of chromospheric variability from multiple satellites, {\it Journal of Geophysical Research}, {\bf 104}, 9995-10005, 1999, DOI:10.1029/1998JA900163.

\bibitem[Viereck et al.(2001)]{Viereck01} Viereck, R., L. Puga, D. McMullin, D. Judge, M. Weber, and W.K. Tobiska, The Mg II index: A proxy for solar EUV, {\it Geophysical Research Letters}, {\bf 4}, 1343–1346, 2001, DOI:10.1029/2000GL012551.

\bibitem[Woods et al.(2000)]{Woods}Woods, T., G. Rottman, J. Harder, G. Lawrence, B. McClintock, G. Kopp, and C. Pankratz, Overview of the EOS SORCE Mission, {\it SPIE Proceedings}, {\bf 4135}, 192, 2000, DOI:10.1117/12.494229.

\bibitem[Yeo et al.(2014)]{Yeo}Yeo, K.L., N.A. Krivova, S.K. Solanki, and K.H. Glassmeier, Reconstruction of total and spectral solar irradiance from 1974 to 2013 based on KPVT, SoHO/MDI, and SDO/HMI observations, {\it Astronomy \& Astrophysics}, {\bf 570}, A85–A85, 2014, DOI:10.1051/0004-6361/201423628.

\end{thebibliography}

%\end{linenumbers}

%\end{document}

%%    If you wish to include your bibliography items in your tex file 
%%    using {thebibliography} as shown below you must out-comment the 
%%    three lines above (insert % at the start of each line) 

%\end{linenumbers}

\end{document}